\documentclass[notitlepage,aps,prl,reprint,superscriptaddress]{revtex4-1}

\renewcommand{\figurename}{FIG.}

\usepackage{qcircuit}
\usepackage{mathtools}
\usepackage[english]{babel}
\usepackage[utf8]{inputenc}
\usepackage{amsmath}
\usepackage{amsfonts}
\usepackage{amssymb}
\usepackage{graphicx}
\usepackage{bbold}
\usepackage{epstopdf}
\usepackage{hyperref}
\renewcommand*{\Re}{\operatorname{Re}} 
\renewcommand*{\Im}{\operatorname{Im}} 
\DeclarePairedDelimiter\bra{\langle}{\rvert}
\DeclarePairedDelimiter\ket{\lvert}{\rangle}

\newcommand{\appropto}{\mathrel{\vcenter{
  \offinterlineskip\halign{\hfil$##$\cr
    \propto\cr\noalign{\kern2pt}\sim\cr\noalign{\kern-2pt}}}}}

\begin{document}
\renewcommand{\figurename}{FIG.}
\renewcommand{\tablename}{TABLE}
\title{Improved readout of qubit-coupled Gottesman-Kitaev-Preskill states}
\author{Jacob Hastrup}
\email{jhast@fysik.dtu.dk}
\author{Ulrik Lund Andersen}

\affiliation{Center for Macroscopic Quantum States (bigQ), Department of Physics, Technical University of Denmark, Building 307, Fysikvej, 2800 Kgs. Lyngby, Denmark}

\begin{abstract}
The Gottesman-Kitaev-Preskill encoding of a qubit in a harmonic oscillator is a promising building block towards fault-tolerant quantum computation. Recently, this encoding was experimentally demonstrated for the first time in trapped-ion and superconducting circuit systems. However, these systems lack some of the Gaussian operations which are critical to efficiently manipulate the encoded qubits. In particular, homodyne detection, which is the preferred method for readout of the encoded qubit, is not readily available, heavily limiting the readout fidelity. Here, we present an alternative read-out strategy designed for qubit-coupled systems. Our method can improve the readout fidelity with several orders of magnitude for such systems and, surprisingly, even surpass the fidelity of homodyne detection in the low squeezing regime. 
\end{abstract}
\date{\today}

\maketitle

\section{Introduction}
Scalable fault-tolerant quantum computation requires physical qubits which can be stored and manipulated with very high fidelity. One approach for realising such high quality qubits is to encode each qubit into a quantum harmonic oscillator. There are several proposals for such encodings, e.g. the cat code \cite{leghtas2013hardware,ofek2016extending}, binomial code \cite{michael2016new,hu2019quantum} and Gottesman-Kitaev-Preskill (GKP) code \cite{gottesman2001encoding,tzitrin2019towards,terhal2020towards,fluhmann2019encoding,campagne2019quantum}. The GKP code has the advantageous property that a universal set of operations can performed using solely Gaussian resources combined with the computational basis states \cite{baragiola2019all,yamasaki2020cost}, and it can be combined with continuous variable cluster states \cite{menicucci2014fault,fukui2018high} or the surface code \cite{vuillot2019quantum,noh2020fault} to achieve fault-tolerance. Furthermore, the GKP code has been shown to outperform other encoding schemes in terms of its efficiency in correcting against loss \cite{albert2018performance,noh2018quantum}, which is the main noise factor in many continuous variable systems. These favourable features have sparked numerous new studies on applying GKP states for optical quantum computing \cite{tzitrin2019towards,larsen2019deterministic,asavanant2019generation}. Still, the generation of GKP states in the optical regime has proven extremely challenging and has so far not been demonstrated experimentally, despite several theoretical proposals \cite{vasconcelos2010all,weigand2018generating,motes2017encoding,su2019conversion,eaton2019non}. However, recently GKP states were generated for the first time in the motional state of a trapped ion \cite{fluhmann2019encoding} and in a microwave cavity field coupled to a superconducting circuit \cite{campagne2019quantum}. These experiments were made possible by the strong coupling between a bosonic mode and an ancillary qubit, enabling non-Gaussian transformation of the bosonic mode. Yet, these experimental platforms lack some of the crucial Gaussian operations that are required for complete manipulation, stabilization and read-out of the encoded GKP qubit \cite{gottesman2001encoding}. Therefore, new methods specifically designed to qubit-coupled systems are required to take full advantage of the GKP encoding in these systems. For example, stabilization has already been demonstrated using the qubit-coupling \cite{campagne2019quantum}, but the lack of homodyne detection severely limits the read-out fidelity \cite{terhal2020towards}. 

Here we propose an improved readout scheme for qubit-coupled GKP states. Our method relies on mapping the logical information of the GKP qubit onto the ancilla qubit state. This is similar to the known method based on phase-estimation \cite{terhal2016encoding}, but by adding an additional interaction between the qubit and the oscillator, we achieve much higher read-out fidelity. For example, for a 10 dB squeezed GKP state our method improves the read-out fidelity from $96.22\%$ with known techniques to $99.98\%$. 

\section{Preliminaries}
We consider a bosonic mode with quadrature operators $\hat{X}$ and $\hat{P}$ satisfying $[\hat{X},\hat{P}] = i$. The code states of the square GKP encoding are defined in the common $+1$ eigenspace of the commuting displacement operators $\hat{D}(\sqrt{2\pi})$ and $\hat{D}(i\sqrt{2\pi})$, where $\hat{D}(\alpha)=e^{\sqrt{2}i\left(-\Re[\alpha]\hat{P} + \Im[\alpha]\hat{X}\right)}$. The computational basis states are then defined as the $\pm1$ eigenstates of $\hat{D}(i\sqrt{\pi}/2)$, which acts as a logical $\hat{Z}$ operator. However, ideal GKP states are unphysical, as they require infinite energy. Instead, the physically relevant basis states are thus only approximate eigenstates of the logical $\hat{Z}$ operator, i.e. $\langle \hat{D}(i\sqrt{\pi}/2)\rangle \approx \pm1$. There are multiple ways of expressing such approximate states. In this work we initially consider the commonly used expression for which the basis states consist of a superposition of multiple squeezed states of width $\Delta$, under a Gaussian envelope of width $\kappa$:
\begin{subequations}
\begin{align}
\ket{\tilde{0}} &\propto \sum_{s\in\mathbb{Z}} e^{-\left(\sqrt{\frac{\pi}{2}}2s\right)^2/\kappa^2}\hat{D}\left(\sqrt{\frac{\pi}{2}}2s\right)\hat{S}_\Delta\ket{\textrm{vac}}\\
\ket{\tilde{1}} &\propto \sum_{s\in\mathbb{Z}} e^{-\left(\sqrt{\frac{\pi}{2}}(2s+1)\right)^2/\kappa^2}\hat{D}\left(\sqrt{\frac{\pi}{2}}(2s+1)\right)\hat{S}_\Delta\ket{\textrm{vac}},    
\end{align}
\label{eq:GKP}
\end{subequations}
where $\ket{\textrm{vac}}$ is the vacuum state and $\hat{S}_\Delta = e^{\frac{i}{2}\ln(\Delta)\left(\hat{X}\hat{P}+\hat{P}\hat{X}\right)}$ is the squeezing operator. The amount of squeezing is often expressed in dB as $\Delta_\textrm{dB} = -10\log_{10}(\Delta^2)$. The approximate code states approach the ideal states for $(\Delta,\kappa^{-1})\rightarrow 0$. It is common to consider the symmetric case where $\Delta=\kappa^{-1}$, but in this paper only $\Delta$ is relevant.

We now consider the problem of how to reliably distinguish between the states $\ket{\tilde{0}}$ and $\ket{\tilde{1}}$ in a physically relevant setting. In particular, we wish to minimize the measurement error probability
\begin{equation}
    p_\textrm{err} = \frac{1}{2}(p(1|0) + p(0|1)),
\end{equation} where $p(x|y)$ is the probability of obtaining measurement outcome $x$ given the input state $y$. Since the approximate states $\ket{\tilde{0}}$ and $\ket{\tilde{1}}$ are not orthogonal, this problem is ultimately bounded by the Helstrom bound:
\begin{equation}
     p_\textrm{err}\geq p_\textrm{err,Helstrom} = \frac{1}{2}\left(1 -\sqrt{1 - |\langle \tilde{0}| \tilde{1}\rangle|^2}\right).
\end{equation}
The Helstrom bound drops very rapidly with decreasing $\Delta$, but is generally not achievable in a physical setting. Instead, homodyne detection is often considered as a practical and efficient read-out method. With this method, the state is measured in the bosonic $\hat{X}$-basis, and the results closer to even multiples of $\sqrt{\pi}$ are considered a 0 while results closer to an odd multiple of $\sqrt{\pi}$ are considered a 1. The measurement error probability for homodyne detection is given by:
\begin{equation}
    p_\textrm{err,homodyne} = \textrm{erfc}\left(\frac{\sqrt{\pi}}{2\Delta}\right) \approx \frac{2}{\pi}\Delta e^{-\frac{1}{4\Delta^2}}, \label{eq:homodyne}
\end{equation}
assuming a negligible overlap between neighbouring squeezed states of the basis states, i.e. $|\bra{\textrm{vac}}\hat{S}_\Delta^\dagger\hat{D}(\sqrt{2\pi})\hat{S}_\Delta\ket{\textrm{vac}}|\approx0$. The exponential term in Eq. \eqref{eq:homodyne} causes the measurement error probability to drop rapidly with decreasing $\Delta$, i.e. homodyne detection is very efficient for highly squeezed states.

However, while homodyne detection can be efficiently implemented in free-space optics, it is less practical for microwave cavities or trapped ions. Instead, these system can couple to an ancilla qubit, e.g. a superconducting transmon qubit for the microwave platform or an internal spin state for the trapped ions, and the state of the ancilla qubit can subsequently be measured. In particular, it is possible to realise a Rabi-type interaction Hamiltonian, $\hat{X}\hat{\sigma}_x$, where $\hat{\sigma}_x$ is the Pauli-$x$ operator of the qubit \cite{fluhmann2018sequential,campagne2019quantum}. The action of this Hamiltonian is sometimes referred to as a conditional displacement, as the bosonic mode gets displaced in a direction depending on the state of the qubit, entangling the qubit and the oscillator. Such interaction can be used to read-out a GKP-qubit using the following simple circuit \cite{terhal2020towards}:
\begin{equation}
\Qcircuit @C=1.2em @R=1.6em {
\lstick{\underset{\textrm{bosonic mode}}{\ket{\psi}_\textrm{GKP}}} & \multigate{1}{U_x\left(i\frac{\sqrt{\pi}}{2}\right)} & \qw  \\
\lstick{\underset{\textrm{qubit}}{\ket{0}}} & \ghost{U_x\left(i\frac{\sqrt{\pi}}{2}\right)} & \meter 
} \label{circuit:simple}
\end{equation}
where 
\begin{equation}
    U_k(\alpha) = \exp\left[i\left(-\Re[\alpha]\hat{P}+\Im[\alpha]\hat{X}\right)\hat{\sigma}_k\right]
\end{equation} 
for $k\in\{x,y,z\}$. The expected measurement outcome of the qubit is $\frac{1}{2}\left(1+\Re\big\langle \hat{D}\left(i\sqrt{\frac{\pi}{2}}\right)\big\rangle\right)$. For ideal GKP basis states for which $\big\langle\hat{D}\left(i\sqrt{\frac{\pi}{2}}\right)\big\rangle=\pm1$ we achieve a perfect read-out. For the approximate states $\ket{\tilde{0}}$ and $\ket{\tilde{1}}$ for which $\big\langle \hat{D}\left(i\sqrt{\frac{\pi}{2}}\right)\big\rangle = \pm e^{-\frac{\pi}{4}\Delta^2}$, the measurement error probability is:
\begin{align}
    p_\textrm{err,simple} = \frac{1}{2}\left(1-e^{-\frac{\pi}{4}\Delta^2}\right) \approx \frac{\pi}{8}\Delta^2. \label{eq:simple}
\end{align}
This scaling is significantly worse than the homodyne strategy of Eq. \eqref{eq:homodyne}. The scaling can be improved by running the circuit multiple times and considering a majority vote, but because of the measurement back-action this strategy has diminishing returns. Additionally, multiple runs of the circuit results in an increased total measurement time during which the state accumulates noise. 

\section{Protocol}
In this work we propose to modify the circuit in \eqref{circuit:simple}, adding an additional Rabi-type interaction of the type $\hat{P}\hat{\sigma}_y$ with interaction strength $\lambda$:
\begin{equation}
\Qcircuit @C=1.2em @R=1.4em {
\lstick{\ket{\psi}_\textrm{GKP}} & \multigate{1}{U_y\left(-\lambda\right)} & \multigate{1}{U_x\left(i\frac{\sqrt{\pi}}{2}\right)} & \qw  \\
\lstick{\ket{0}} & \ghost{U_y\left(-\lambda\right)} & \ghost{U_x\left(i\frac{\sqrt{\pi}}{2}\right)} & \meter 
} \label{circuit:new}
\end{equation}
For $|\lambda|\ll 1$, the measurement error probability of this circuit is given by:
\begin{align}
    p_\textrm{err,improved} &= \frac{1}{2}\left(1-e^{-\frac{\pi}{4}\Delta^2}\left(e^{-\frac{\lambda^2}{\Delta^2}}+\sin(\sqrt{\pi}\lambda)\right)\right),  \label{eq:improved}
\end{align}
which reduces to that of Eq. \eqref{eq:simple} for $\lambda=0$ as expected. However, for $\lambda\neq 0$ it is possible to achieve a better scaling. The minimum is achieved for $\lambda$ satisfying $\frac{2\lambda}{\Delta^2}e^{-\lambda^2/\Delta^2}=\sqrt{\pi}\cos(\sqrt{\pi}\lambda)$, which for small $\Delta$ is approximately at $\lambda=\sqrt{\pi}\Delta^2/2$. Inserting this into Eq. \eqref{eq:improved} and expanding to lowest order in $\Delta$ we get:
\begin{equation}
    p_\textrm{err,improved} \approx \underbrace{\frac{5\pi^3}{384}}_{\sim 0.4}\Delta^6,
\end{equation}
\begin{figure}[t!]
    \centering
    \includegraphics{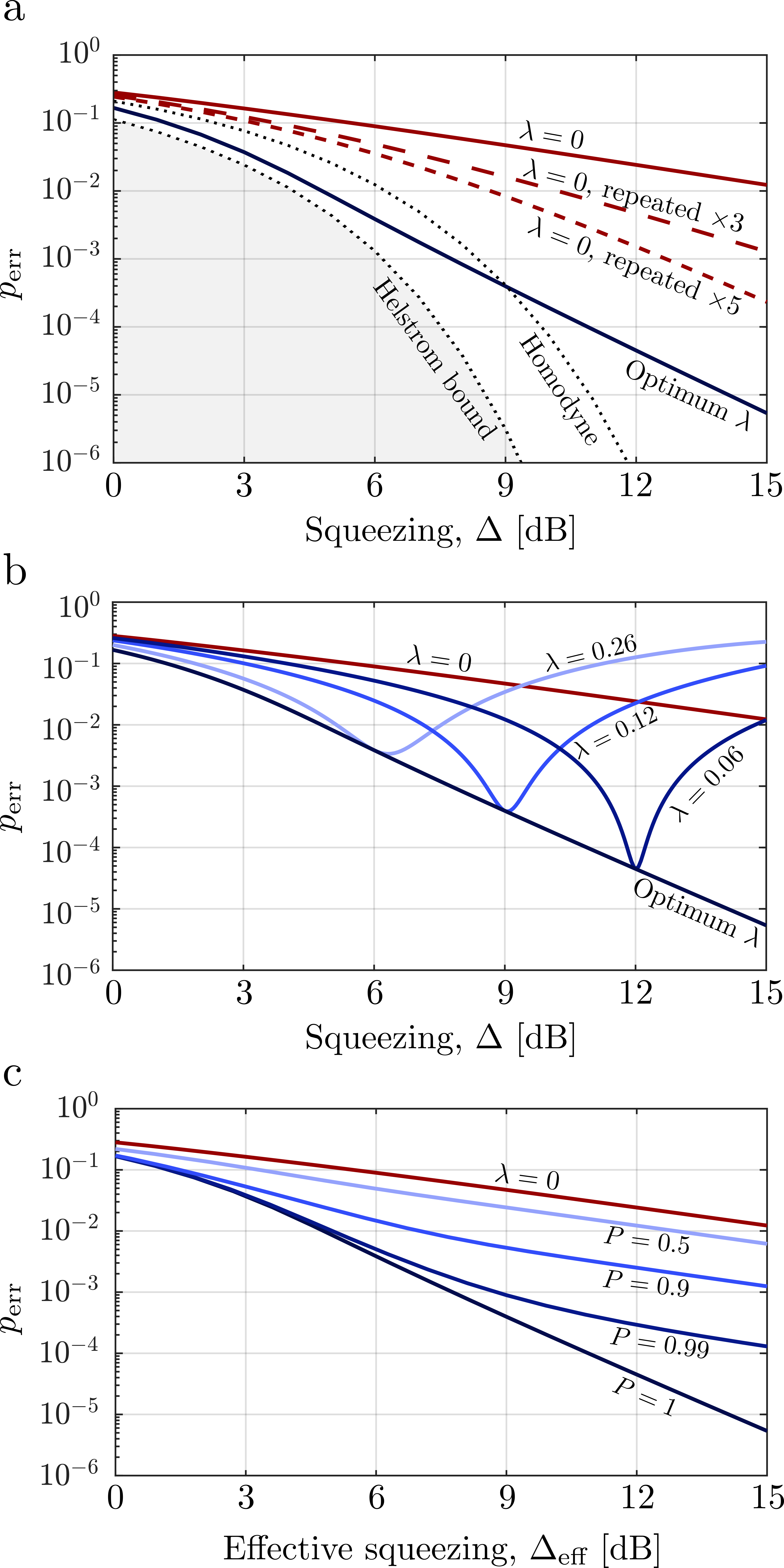}
    \caption{(a): Measurement error probability, $p_\textrm{err}$, for various measurement strategies. The red $\lambda=0$ lines correspond to circuit \eqref{circuit:simple}, while the blue line corresponds to circuit \eqref{circuit:new} with the interaction parameter $\lambda$ chosen to minimize $p_\textrm{err}$. (b): Performance for fixed $\lambda$ as a function of the input squeezing. For large amounts of squeezing the optimal performance is only achieved in a narrow range, requiring good knowledge of the input state. (c): Performance for mixed states generated by applying the Gaussian displacement channel, Eq. \eqref{eq:mixed}, to the pure input states of Eq. \eqref{eq:GKP}. For such states, the purity, $P$, heavily impacts the performance of the protocol, although the performance is always improved compared to the simple circuit.}
    \label{fig:result}
\end{figure}
i.e. a significantly better scaling than \eqref{eq:simple}. The measurement error probabilities of the different methods are compared in Fig. \ref{fig:result}a. The blue curve shows the result of circuit \eqref{circuit:new}, with the optimum $\lambda$ chosen for each point. We see a clear improvement over the simple circuit in \eqref{circuit:simple}, i.e. for $\lambda=0$, even when using multiple runs of the simple circuit. For a squeezing of less than 9 dB the modified circuit even outperforms homodyne detection. We found that using circuit \eqref{circuit:new} we could not further improve the performance using multiple rounds and majority voting. This is because the measurement back-action upon getting the wrong measurement heavily modifies the input state, making subsequent measurement rounds useless.
One important thing to note is, that the optimum interaction parameter, $\lambda$, depends on the quality, or $\Delta$, of the input GKP state. This is different from the homodyne measurement strategy or the simple circuit, both of which are constructed independently on the quality of the input state. Therefore, it is important to calibrate the modified measurement circuit, i.e. tuning $\lambda$, according to the squeezing of the input state. Fig. \ref{fig:result}b shows the performance when fixing $\lambda$ at different values. For large amounts of squeezing we see that the circuit performs optimally only for input states in a narrow region. In a practical setting it might be difficult to consistently fix the squeezing level of the state to be measured, as it could depend on previous operations of the state. Therefore, the average measurement error probability will likely be higher than what is predicted by Eq. $\eqref{eq:improved}$. However, from Fig. \ref{fig:result}b we see that the results are generally improved compared to the simple circuit for a wide range of $\Delta$.

So far we have considered only the states of Eq. \eqref{eq:GKP}. However, these states might not necessarily be physically realistic as, for example, they are pure. Instead, we can construct more general mixed GKP states by applying a Gaussian displacement channel of strength $\sigma$ to the pure states of Eq. \eqref{eq:GKP}:
\begin{equation}
    \rho_\mu = \frac{1}{\pi\sigma^2}\int d^2\alpha e^{-\frac{|\alpha|^2}{\sigma^2}}\hat{D}(\alpha)\ket{\tilde{\mu}}\bra{\tilde{\mu}}\hat{D}^\dagger(\alpha), \label{eq:mixed}
\end{equation}
where $\rho_\mu$ is the density matrix of the output state and $\mu\in\{0,1\}$. The performance of the simple circuit \eqref{circuit:simple} does not depend on the exact form of the input state but only on the expectation value $\big\langle\hat{D}\left(i\sqrt{\frac{\pi}{2}}\right)\big\rangle$. In fact, one can use a similar expectation value to define an effective squeezing parameter $\Delta_\textrm{eff}$ as \cite{duivenvoorden2017single}:
\begin{equation}
    \Delta_\textrm{eff}=\sqrt{\frac{1}{2\pi}\ln\left(\frac{1}{|\langle\hat{D}(i\sqrt{2\pi})\rangle|^2}\right)},
\end{equation}
allowing us to describe the amount of squeezing in an arbitrary state. For the states of \eqref{eq:GKP} we simply have $\Delta_\textrm{eff}=\Delta$. For the mixed state of \eqref{eq:mixed} we have $\Delta_\textrm{eff}=\sqrt{\Delta^2 + 2\sigma^2}$. By tuning $\Delta$ and $\sigma$ we can thus now construct GKP states of arbitrary purity, $P=\textrm{Tr}(\rho^2)$, and effective squeezing. Fig. \ref{fig:result}c shows the performance of the circuit for states of different purity. We see that the performance degrades for mixed states, although we still obtain superior behavior compared to the simple circuit. In the literature, GKP states are commonly only quantified in terms of their squeezing level, with the purity being less relevant as it plays no role for e.g. homodyne detection. It is therefore unclear what levels can be expected in experimental setting, which will also likely vary between platforms. Note that the mixed states were constructed in one particular way in this paper, e.g. by combining Eqs. \eqref{eq:GKP} and \eqref{eq:mixed}. The purity alone might therefore not accurately describe performance of the protocol for other states. Still, the results of Fig. \ref{fig:result}c indicates that high quality states with features beyond just the squeezing are required to take full advantage of the improved measurement scheme. 

\section{Conclusion}
We have presented a protocol for efficient read-out of a GKP state in a qubit-coupled oscillator. Our protocol reduces the measurement error rate from a $\Delta^2$-scaling with previously known methods to a $\Delta^6$-scaling, enabling low error rates in the absence of homodyne detection. Our protocol is sensitive to the exact form of the input state, with a reduced performance for mixed states. However, our results demonstrate that homodyne detection might not be crucial to efficiently utilize the GKP encoding, e.g. in microwave cavities or trapped ions.

\begin{acknowledgements}
\section*{Acknowledgements}
This project was supported by the Danish National Research Foundation through the Center of Excellence for Macroscopic Quantum States (bigQ).
\end{acknowledgements}

\section*{References}
\bibliography{References} 

\end{document}